\documentclass[twoside,twocolumn,american]{revtex4-1}
\usepackage{amsmath,amssymb}
\usepackage{fourier}

\usepackage[latin9]{inputenc}
\usepackage{color}
\usepackage{graphicx}

\begin{document}

\title{Tailoring steep density profile with unstable points}
\author{Shun Ogawa}
\email{shun.ogawa@riken.jp}
\affiliation{Laboratory for Neural Computation and Adaptation,\\ RIKEN Center for Brain
Science, 2-1 Hirosawa Wako Saitama 351-0198, Japan}
\author{Xavier Leoncini}
\affiliation{Aix Marseille Univ, Universit\'e de Toulon, CNRS, CPT, Marseille, France}
\email{Xavier.Leoncini@cpt.univ-mrs.fr}
\author{Alexei Vasiliev}
\affiliation{Space Research Institute, Profsoyuznaya 84/32, Moscow 117997, Russia}
\author{Xavier Garbet}
\affiliation{CEA, IRFM, F-13108 St. Paul-lez-Durance cedex, France}

\begin{abstract}

The mesoscopic properties of a plasma in a cylindrical magnetic field
are investigated from the view point of test-particle dynamics. When
the system has enough time and spatial symmetries, a Hamiltonian of
a test particle is completely integrable and can be reduced to a single
degree of freedom Hamiltonian for each initial state. The reduced
Hamiltonian sometimes has unstable fixed points (saddle points) and
associated separatrices. To choose among available
dynamically compatible equilibrium states of the one particle density
function of these systems we use a maximum entropy principle and
discuss how the unstable fixed points affect the density profile or
a local pressure gradient, and are able to create a steep profile
that improves plasma confinement.
\end{abstract}


\maketitle

Being able to sustain a steep density profile in hot magnetized plasma
is one of the major key points to achieve magnetically confined fusion
devices. These steep profiles are typically associated with the emergence
in the plasma of so-called internal transport barriers (ITB) \cite{Wolf2003,Connor2004}.
Both the creation and study of these barriers have generated numerous
investigations mostly numerical using either a fluid, or magnetic
field or kinetic perspective or combining some of these. In this paper
starting from the direct study of particle motion, we propose a simple
mechanism to set up a steep profile which may not have been fully
considered yet. Indeed, charged particle motion in a non-uniform magnetic
field \cite{Alfven1940,Northrop1961,Littlejohn1981,Boozer2004,Cary2009,Brizard2007,Jackson}
is one of main classical issues of physics of plasmas in space or
in fusion reactors. To tackle this problem the guiding center \cite{Cary2009}
and the gyrokinetic \cite{Brizard2007} theories are developed to
trace the particle's slower motion by averaging the faster cyclotron
motion. These reductions suppress computational cost and they are
widely used to simulate the magnetically confined plasmas in fusion
reactors \cite{Boozer2004}. These reduction theories assume existence
of an invariant or an adiabatic invariant of motion associated with
the magnetic moment. Meanwhile, this assumption does not always hold
true. Then, recently, studies on full particle orbits without any
reductions are done to look into phenomena ignored by these reductions
and to interpolate the guiding center orbit. There exists a case that
a guiding center trajectory and a full trajectory are completely different
\cite{Pfeffrle2015}. Further, it is found that the assumption of
the invariant magnetic moment breaks \cite{Cambon2014,Ogawa2016}
due to the chaotic motion of the test particles.

Let us quickly review the single particle motion and adiabatic chaos.
We consider a model of charged particle moving in a non-uniform cylindrical
magnetic field $\mathbf{B}(r)=\nabla\wedge\mathbf{A}_{0}(r)$. The
vector potential $\boldmath{A}_{0}(r)$ is given by 
\begin{equation}
\begin{split}\mathbf{A}_{0}(r)=\frac{B_{0}r}{2}\mathbf{e}_{\theta}-B_{0}F(r)\mathbf{e}_{z},\quad F(r)=\int_{0}^{r}\frac{rdr}{R_{{\rm per}}q(r)},\end{split}
\label{eq:model}
\end{equation}
where the cylinder is parametrized with the coordinate $(r,\theta,z)$,
$B_{0}$ is strength of the magnetic field, $z$ has $2\pi R_{{\rm per}}$-periodicity,
$\mathbf{e}_{\theta}$ and $\mathbf{e}_{z}$ are basic units for each
direction, and $q(r)$ is a winding number called a safety factor
of magnetic field lines. The Hamiltonian of the particle $H=\|\mathbf{v}\|^{2}/2$,
where $\mathbf{v}$ denotes particle's velocity, has three constants
of motion, the energy, the angular momentum, and the momentum, associated
with time, rotational, and translational symmetry of the system respectively,
so that the Hamiltonian $H$ on the six dimensional phase space is
reduced into the single-degree-of-freedom Hamiltonian on the two dimensional
phase space, $(r,p_{r})$-plane \cite{Cambon2014,Ogawa2016}, 
\begin{equation}
\begin{split} & H_{{\rm eff}}(r,p_{r})=p_{r}^{2}/2+V_{{\rm eff}}(r),\\
 & V_{{\rm eff}}(r)=\frac{v_{\theta}^{2}+v_{z}^{2}}{2}=\frac{\left(p_{\theta}r^{-1}-B_{0}r/2\right)^{2}}{2}+\frac{(p_{z}+B_{0}F(r))^{2}}{2}.
\end{split}
\label{eq:hamiltonian}
\end{equation}
where $p_{i}$ stands for the conjugate momentum for $i=r$, $z$,
and $\theta$ respectively, and where $v_{\theta}=r\dot{\theta}$
and $v_{z}=\dot{z}$. The upper dot denotes $d/dt$. Invariants $p_{z}$
and $p_{\theta}$ are fixed by the initial condition
of the particle in the 6-dimensional phase space. Appropriately
setting the safety factor $q$ and choosing initial condition, we
can find an unstable fixed point in $(r,p_{r})$ phase plane, which
can induce the adiabatic chaos \cite{Neishtadt1986,Neishtadt1987,Tennyson1986,Cary1986,Leoncini2009}
when the weak magnetic perturbation or the curvature effect added
to the flat torus (cylinder) exist\cite{Ogawa2016,Cambon2014}.

This Letter aims to exhibit one possibility that the unstable fixed
point inducing chaotic motion modifies mesoscopic properties of plasmas,
local density and pressure gradients which are believed to be associated
with the internal transport barriers (ITBs) \cite{Wolf2003,Connor2004}
a feature missed by gyrokinetics, or a pure magneto-hydrodynamic approach.
For this purpose, we shall compute an equilibrium
radial density function $\rho(r)$ from stationary
kinetic distribution. When neglecting the feedback on the fields of
the motion of the particles governed by the Hamiltonian (\ref{eq:hamiltonian})
computing a stationary state of an ensemble of particles, i.e. a stationary
one-particle density function, resumes to find a density function
$f_{0}$ on the phase space, which commutes with Hamiltonian \eqref{eq:hamiltonian}.
Since the motion is integrable, these solutions correspond after a
local change to action-angle variables to functions depending only
on the actions of the Hamiltonian with a uniform distribution of the
associated angles (see for instance for a similar situation \cite{leoncini09}). As a
consequence, there exist infinitely many steady states. In
order to choose one, we may assume a vanishing collisionality, and
consider that the one maximizing the information entropy under suitable
constraint conditions is picked out \cite{Zubarev1996,Rocha2005}.
In principle, we should consider a Vlasov-Maxwell system consisting
of the collisionless Boltzmann equation describing a temporal evolution
of single particle density functions of ions and electrons, coupled
with the Maxwell equation determining a self-consistent electro-magnetic
field \cite{Vlasov1938,Pitaevskii1981}, when neglecting
the self-consistency we notably neglect electrons (their presence
insures a neutralizing background, and the current to get the right
poloidal component of magnetic field), inter-particle interactions,
radiation from moving charged particles and back reaction from the
electro-magnetic field. As such we are looking for
a steady state of a truncated Vlasov-Maxwell system with an ion moving
in a static magnetic field. In this setting we qualitatively
discuss which kind of vector potentials $F(r)$ or $q$-profiles are
likely to bring about the unstable fixed point for the test particle
motion. Then, we look into the effect of the unstable fixed point
for the density profile and the local-pressure gradient. We shall
end this Letter by remarking the relation between the steep density
profile (particle's ITBs) and the magnetic ITBs~\cite{Balescu1998,Constantinescu2012}.

Among the different stationary distributions, let
us now compute the general form of $f_{0}$ which maximizes the information
entropy (also called a density of the Boltzmann Gibbs. The Boltzmann's
constant $k_{{\rm B}}$ is set as unity. We thus have
to maximize the functional 
\begin{equation}
\mathcal{S}[f]=-\iint_{\mu}f\ln fd^{3}\mathbf{p}d^{3}\mathbf{q},
\end{equation}
subject to the normalization condition (conservation of the number
of particles) and energy, momentum, and angular momentum conservations,
which are respectively 
\begin{equation}
\begin{split}\mathcal{N}[f]=\iint_{\mu}fd^{3}\mathbf{p}d^{3}\mathbf{q},\quad\mathcal{E}[f]=\iint_{\mu}H_{{\rm eff}}fd^{3}\mathbf{p}d^{3}\mathbf{q},\\
\mathcal{P}[f]=\iint_{\mu}p_{z}fd^{3}\mathbf{p}d^{3}\mathbf{q},\quad\mathcal{L}[f]=\iint_{\mu}p_{\theta}fd^{3}\mathbf{p}d^{3}\mathbf{q},
\end{split}
\end{equation}
where the integral $\iint_{\mu}\bullet d^{3}\mathbf{p}d^{3}\mathbf{q}$
means average over the six dimensional single particle phase space,
noted here $\mu$-space. The solution to this variational problem
is 
\begin{equation}
\begin{split}f_{0} & =e^{-\beta H_{{\rm eff}}-\gamma_{1}-\gamma_{\theta}p_{\theta}-\gamma_{z}p_{z}}\end{split}
\label{eq:mu-density}
\end{equation}
where $\beta$, $\gamma_{1}$, $\gamma_{\theta}$, and $\gamma_{z}$
are the Lagrangian multipliers, associated with energy conservation,
normalization, momentum and angular momentum conditions respectively.
The parameter $\beta$ corresponds to the thermodynamical temperature
as 
\begin{equation}
T_{{\rm th}}^{-1}\equiv\beta=\delta\mathcal{S}/\delta\mathcal{E}\:,\label{eq:Temperature}
\end{equation}
 and it can be safely assumed positive. It should be noted that $-\gamma_{\theta}$
and $-\gamma_{z}$ are proportional to the ensemble averages of $v_{\theta}$
and $v_{z}$ respectively. In the literature is has been admitted
that when an ITB exists plasma rotation exists. We then expect that
in such state the averages of $v_{\theta}$ and $v_{z}$ are not 0
and so are $\gamma_{\theta}$ and $\gamma_{z}$.
The spatial density function $n(\boldmath{q})$ is deduced from this
result as 
\begin{equation}
n(\mathbf{q})\equiv\int f_{0}d^{3}\mathbf{p}=\int f_{0}r^{-1}dp_{\theta}dp_{z}dp_{r}.
\end{equation}
Thus, the density $n(r)d^{3}\boldmath{q}$ is proportional to 
\begin{equation}
n(\mathbf{q})rdrd\theta dz\propto e^{\left(\frac{\gamma_{\theta}}{2}\left(-B_{0}-\frac{\gamma_{\theta}}{\beta}\right)r^{2}+\gamma_{z}B_{0}F(r)\right)}rdrd\theta dz,
\end{equation}
and is independent of $\theta$ and $z$. We then obtain a radial
density function $\rho(r)$ given by 
\begin{equation}
\rho(r)=\frac{\iint n(\mathbf{q})rd\theta dz}{\iint rd\theta dz}=\frac{1}{4\pi^{2}rR_{{\rm per}}}\iint n(\mathbf{q})rd\theta dz,
\end{equation}
as 
\begin{equation}
\begin{split}\rho(r) & =\frac{\exp\left(-ar^{2}-bF(r)\right)}{\int_{0}^{{\color{red}\infty}}\exp\left(-ar^{2}-bF(r)\right)dr},\\
a & =\frac{\gamma_{\theta}}{2}\left(B_{0}-\frac{\gamma_{\theta}}{\beta}\right),\quad b=-\gamma_{z}B_{0}.
\end{split}
\label{eq:density}
\end{equation}
We can notice from Eq.(\ref{eq:density}) that the
equilibrium profile is not flat as soon as $\gamma_{\theta}$ is not
zero and that it depends on the poloidal magnetic field configuration
when $\gamma_{z}\ne0$. Given the definitions, this means as soon
as the plasma moves the profiles are not flat. Since we are considering
an equilibrium configuration, we may as well end up with a non-flat
temperature profile, but here we have to consider the local radial
kinetic temperature of the particles rather than the thermodynamic
one (\ref{eq:Temperature}), so this would correspond to the average
of the energy at constant radius. In the same spirit as for the density
we can compute the spatial energy density function $\varepsilon(\boldmath{q})$
is deduced from this result as 
\begin{equation}
\varepsilon(\mathbf{q})\equiv\int f_{0}H_{{\rm eff}}d^{3}\mathbf{p}=\int f_{0}H_{{\rm eff}}r^{-1}dp_{\theta}dp_{z}dp_{r}.
\end{equation}
Thus, we notice that 
\begin{equation}
\varepsilon(\mathbf{q})=-\frac{\partial n(\mathbf{q})}{\partial\beta}\label{eq:11_bis}
\end{equation}
so the kinetic temperature profile $T(r)$ is proportional to $\rho(r)$.
In the same spirit it should be noted that the local pressure $P(r)$
is proportional to the radial density $\rho(r)$, because we assume
the equation of state $P(r)=N\rho(r)T_{{\rm th}}$ holds locally true,
where $N$ is the number of particles and here we
consider the equilibrium temperature $T_{{\rm th}}$.

We now move on and consider how the existence of the unstable fixed
points with relevant energy level affects the obtained equilibrium
density profile. For this purpose we have to discuss how the safety
factor is chosen, in other words which function $F$ in Eq. \eqref{eq:model}
leads to the emergence of ``practical'' unstable points in the effective
potential $V_{{\rm eff}}$. Indeed, as a first point to pin out, if
the amplitude of $F$ is large, we can expect that the term $v_{z}^{2}/2=(p_{z}+F)^{2}/2$
in the Hamiltonian \eqref{eq:hamiltonian} becomes also large, then
the unstable points appear in the phase space at so high energy level
that they become physically irrelevant. Therefore, the amplitude of
$F$ should be small.

\begin{figure}[tb]
\centering{} \includegraphics[width=8cm]{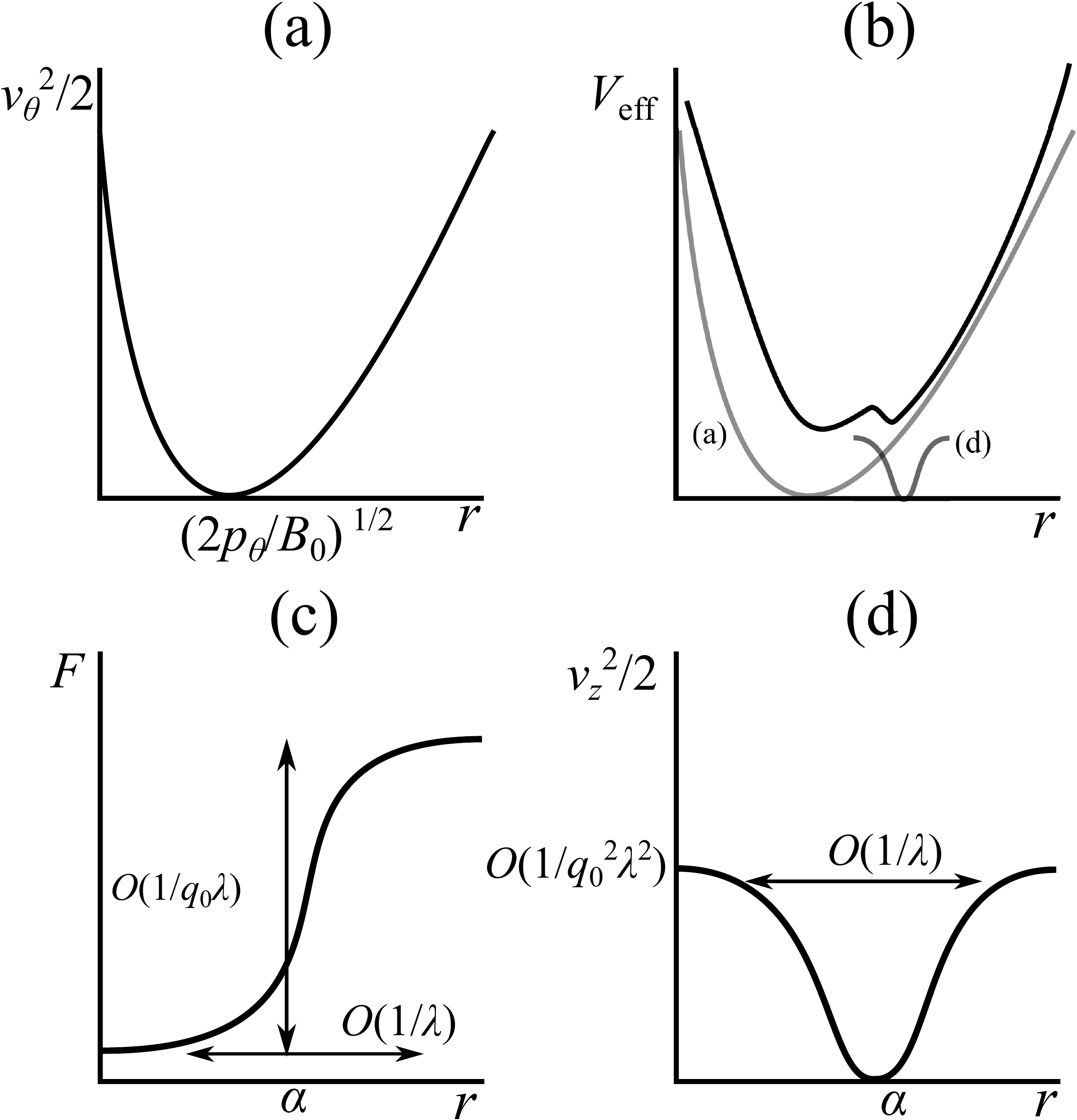} \caption{ Schematic picture showing how the saddle point appears around the
bottom of safety factors. A Panel (a) represents a part of $v_{\theta}^{2}/2=(p_{\theta}/r-B_{0}r/2)^{2}/2$,
(b) $V_{{\rm eff}}$, (c) $F(r)$, and (d) $v_{z}^{2}/2=(p_{z}+F(r))^{2}/2$.
In the panel (d), we consider that there exists an $r$ such that
$p_{z}+F(r)=0$. The parameters $q_{0}$, $\lambda$, and $\alpha$
are associated with Eq. \eqref{eq:q}. }
\label{fig:schematic6} 
\end{figure}

Moreover if the variations of $F(r)$ are smooth and ``gentle''
with $r$, so does again $(p_{z}+F)^{2}/2$ in $V_{{\rm eff}}$, then
$V_{{\rm eff}}$ has only one minimum point that is essentially governed
by the term $(p_{\theta}/r-B_{0}r/2)^{2}/2$. Thus, enough concavity
of $v_{z}^{2}/2$ near but not at the minimum point of $(p_{\theta}/r-B_{0}r/2)^{2}/2$,
$r=\sqrt{2p_{\theta}/B_{0}}$ is necessary so that $V_{{\rm eff}}$
has unstable points. These considerations are illustrated in Fig.~\ref{fig:schematic6}.
In the panel (d), we assume that there exists an $r$ such that $p_{z}+F(r)=0$.
We stress out as well that if $|p_{z}|$ is sufficiently large, it
is also possible to create an unstable point, but then again the energy
level is so high that it is irrelevant for the mesoscopic profiles
in considered plasmas. 

\begin{figure}[t]
\centering{} \includegraphics[width=8cm]{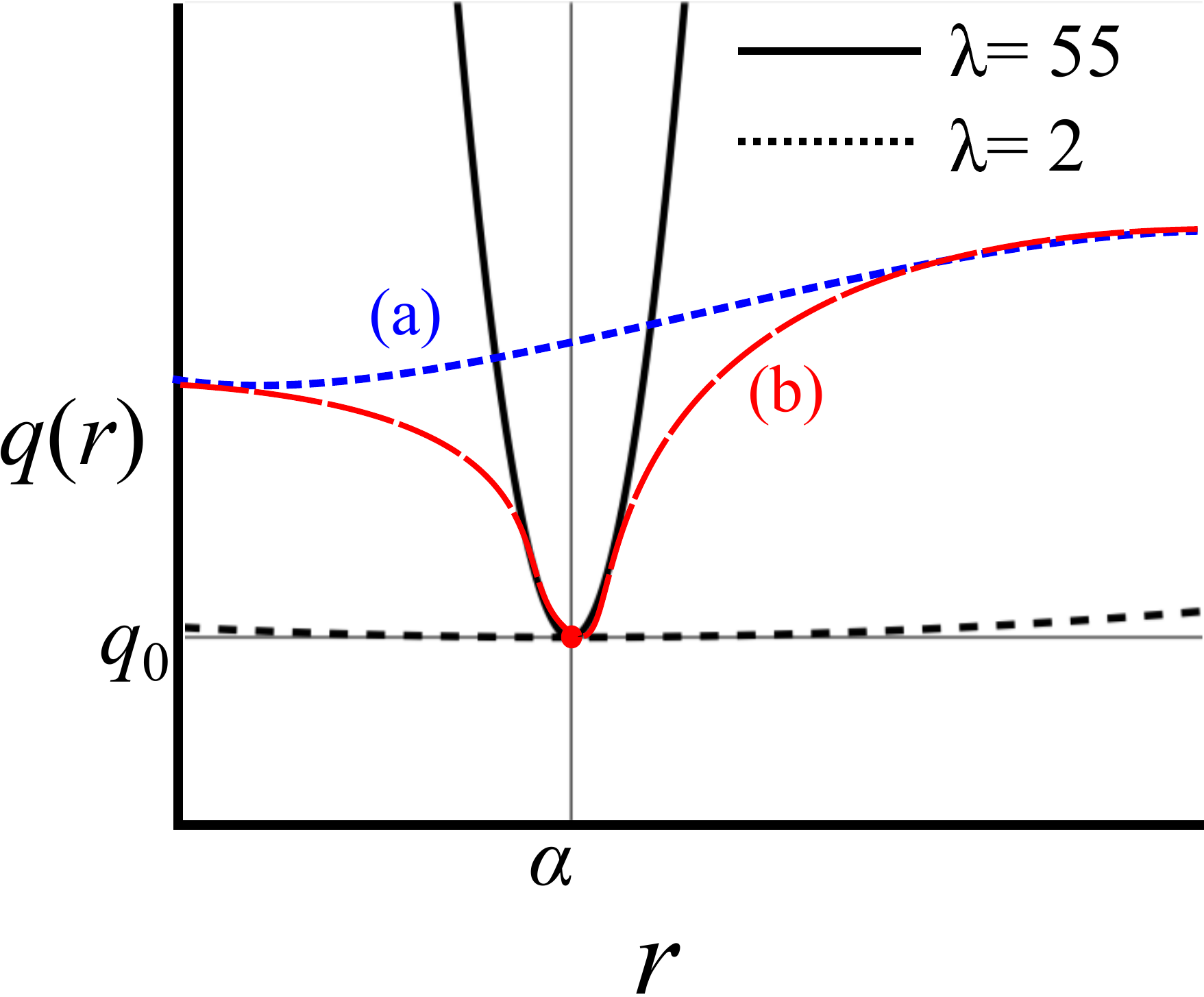} \caption{ Schematic picture exhibiting how to create unstable fixed point by
modifying $q$-profile. The solid curve represents the given $q$-profile.
Dotted and bold curves are the graph of $q_{0}(1+\lambda^{2}(r-\alpha)^{2})$
for $q_{0}=0.12$, $\alpha=\sqrt{0.18}$ and $\lambda=2,55$ respectively.
The $q(r)$ with $\lambda=55$ induces the unstable fixed points,
so does not one with $\lambda=2$. (a) and (b) correspond respectively
to the $q$-profiles without and with unstable fixed points. }
\label{fig:schematic10} 
\end{figure}

Given the obtained density profile (\ref{eq:density}),
we can notice that a sudden fast variation of the function $F$, will
lead to strong variations of the profile, as long as $r$ is not too
large, for instance a step like profile should translate in a steep
profile. With this in mind, since this effect is present if $\gamma_{z},$
related to the average velocity along the cylinder axis, we may expect
that in the context of magnetic fusion with machines with large aspect
ratios the presence of zonal flows along the toroidal direction is
important to increase confinement. Going back to our simple model,
since the safety factor $q(r)$ can be directly associated with the
function $F(r)$ at the origin of the unstable fixed points, and $q(r)$
is a crucial parameter for the operation of magnetized fusion machine,
let us discuss more how the constraints discussed previously translate
on the $q$-profile. For instance let us consider a situation with
a non-monotonous profile such that $q(r)$ has a minimum $q_{0}$
at $r=\alpha$ and the spatial scale is characterized with $\lambda$.
Then locally $q(r)$ can be expressed as 
\begin{equation}
q(r)=q_{0}\left[1+\lambda^{2}(r-\alpha)^{2}\right],\quad r\sim\alpha.\label{eq:q}
\end{equation}
Recalling Eq.~\eqref{eq:model}, the function $F$ is scaled as $q_{0}^{-1}\lambda^{-1}$,
and $v_{z}^{2}/2=(p_{z}+F)^{2}/2\sim q_{0}^{-2}\lambda^{-2}$, so
that this provides a typical energy level of the particles located
near a separatrix. It should be noted that the width of the well of
$v_{z}^{2}/2$ scales as $\lambda^{-1}$ (see Fig.~\ref{fig:schematic6}).
For a fixed value of $q_{0}$, a large value of $\lambda$ creates
unstable fixed points with relevant energy levels for the particle
whose angular momentum is $p_{\theta}\sim B_{0}\alpha^{2}/2$. As
$\lambda$ gets to be larger, the number of the particles with unstable
fixed point increase. This is because, roughly speaking, the energy
levels of unstable points get to be lower, and the one-particle density
\eqref{eq:mu-density} is proportional to $e^{-\beta H_{{\rm eff}}}$.
As a consequence of these considerations we illustrate on Fig.~\ref{fig:schematic10}
how to adjust a given $q$-profile in order to create unstable fixed
points whose location is $r\sim\alpha$. One can set up unstable fixed
point around $r\sim\alpha$ by modifying the $q$-profile so that
it has concavity around $r=\alpha$.

\begin{figure}[t]
\begin{centering}
\includegraphics[width=8cm]{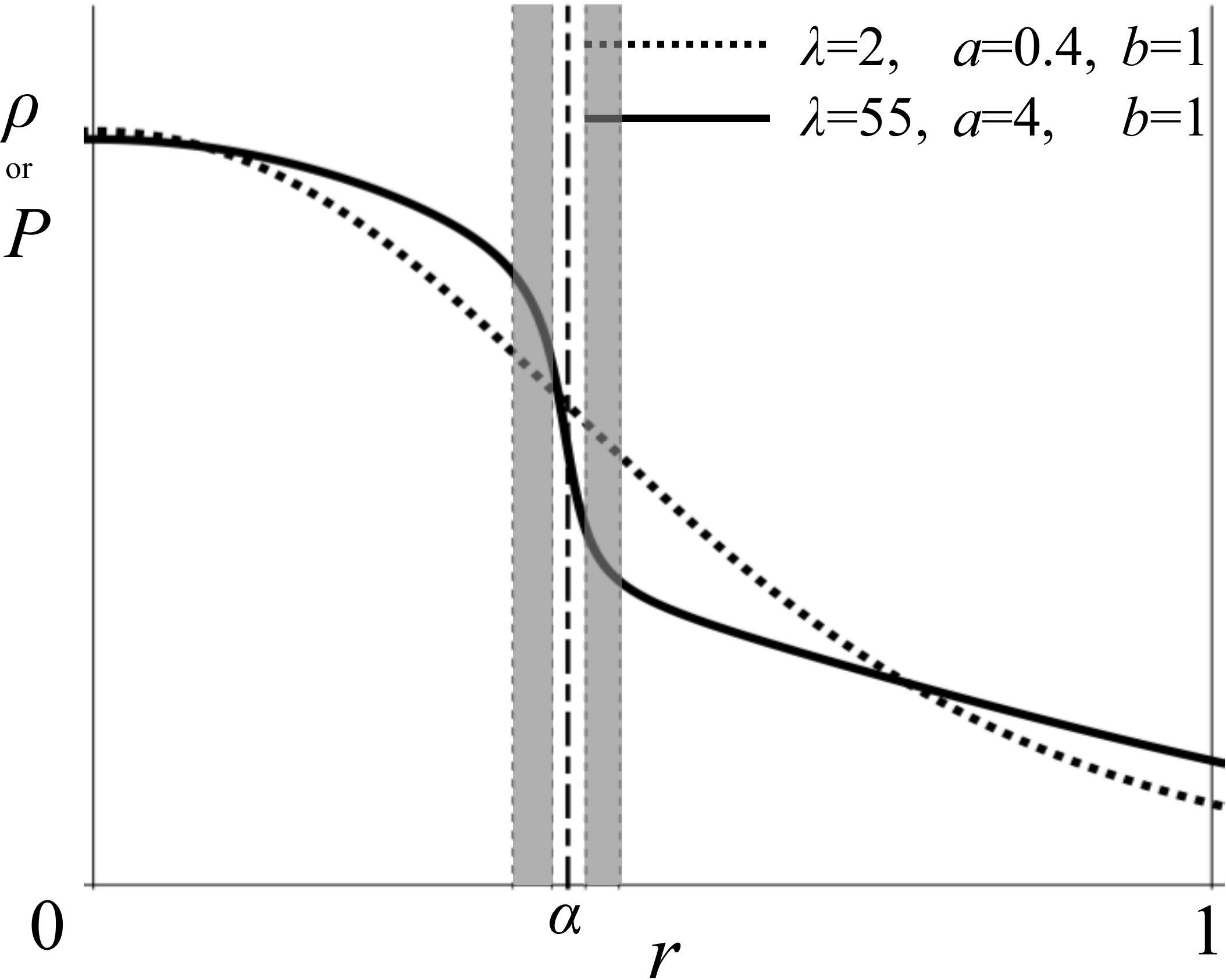} 
\par\end{centering}
\caption{(Color online) The solid and dotted curves express the density profiles
with $q$-profiles with unstable points (shaded region) and without
unstable points respectively. The parameters in a local $q$-profile
{[}Eq. \eqref{eq:q}{]} are determined as $q_{0}=0.12$, and $\alpha=\sqrt{0.18}\simeq0.4243$,
and $\lambda=55$ for the solid curve and $\lambda=2$ for the dotted
one. In both profiles, $\gamma_{z}$ and $\beta$ are same, but $\gamma_{\theta}$
is different. }
\label{fig:density} 
\end{figure}

We then have a form of density profile \eqref{eq:density} and a condition
for $q$-profile exhibiting unstable fixed points. We next consider
where the unstable points appear, and we shall exhibit that the emergence
of unstable fixed points induces the presence of a local steep profile
in their vicinity, \textit{i.e.} their radial positions are inducing
the existence of locally strong density gradients. For this purpose
we simply consider the $q$-profile given by Eq.~\eqref{eq:q} with
parameters $q_{0}=0.12$, $\lambda=55$, and $\alpha=\sqrt{0.18}\simeq0.4243$,
in Figs.~\ref{fig:density}. When including perturbations, we point
out that the adiabatic chaos due to separatrix crossing in this magnetic
field has been discussed in Ref.~\cite{Ogawa2016}. The results are
displayed in Fig.~\ref{fig:density}, where two density profiles
\eqref{eq:density} obtained for two $q$-profiles with and without
unstable fixed points are shown. The parameter $a$ is changed so
that they have same density in the center of cylinder. We note $a$
can be changed keeping the thermodynamical temperature $\beta^{-1}$
and changing the average of angle velocity $v_{\theta}$. We find
the steep region which corresponds to the local steep density gradient
around $r=\alpha$ on which $q(r)$ satisfies $q'(r)=0$ for the $q$-profile
with unstable points. 
Going further on, reasoning directly with a $q$-profile
may lead to some physical problems. For a given magnetic configuration
inspired from a tokamak, the toroidal component of the magnetic field
is generated by the current within the plasma. In the large aspect
ratio (cylindrical) limit, this readily gives a $r$ dependence of
the current 
\begin{equation}
j(r)=\frac{1}{r}\frac{\partial}{\partial r}\left(r\frac{\partial F}{\partial r}\right)\:.\label{eq:Current_function_of_F}
\end{equation}
 When looking at the current profile given by the $q$-profile giving
rise to Fig.~\ref{fig:density}, we end up with two different region
one, with a negative current for $r$ sufficiently large, and one
with a positive one for small $r$. This is likely to be physically
not realistic. Eventhough it may have appeared as possible, the actual stability of these configuration
is doubtful, see for instance \cite{Huysmans01,Hammett03,Pozdnyakov05} .In order to generate a profile that could be more physically
relevant, we may have again a look at Eq.~(\ref{eq:density}). We
can notice that a ``steep'' variation of the function $F(r)$ will
likely trigger a steepness in the density profile. Taking into account
Eq.~(\ref{eq:Current_function_of_F}), we can look for profiles that
can achieve this variation while keeping a positive current. Given
the form of Eq.~(\ref{eq:Current_function_of_F}), it appears as
simpler to look for just one Bessel function of the first kind $F=F_{0}J_{\nu}(\lambda r)$
and to create only one separatrix, we consider the case when no plasma
current is present, then the minimum of the effective potential (\ref{eq:hamiltonian})
is located at $r_{0}$ such that $p_{\theta}=B_{0}r_{0}^{2}/2$. 
\begin{figure}[t]
\begin{centering}
\includegraphics[width=8cm]{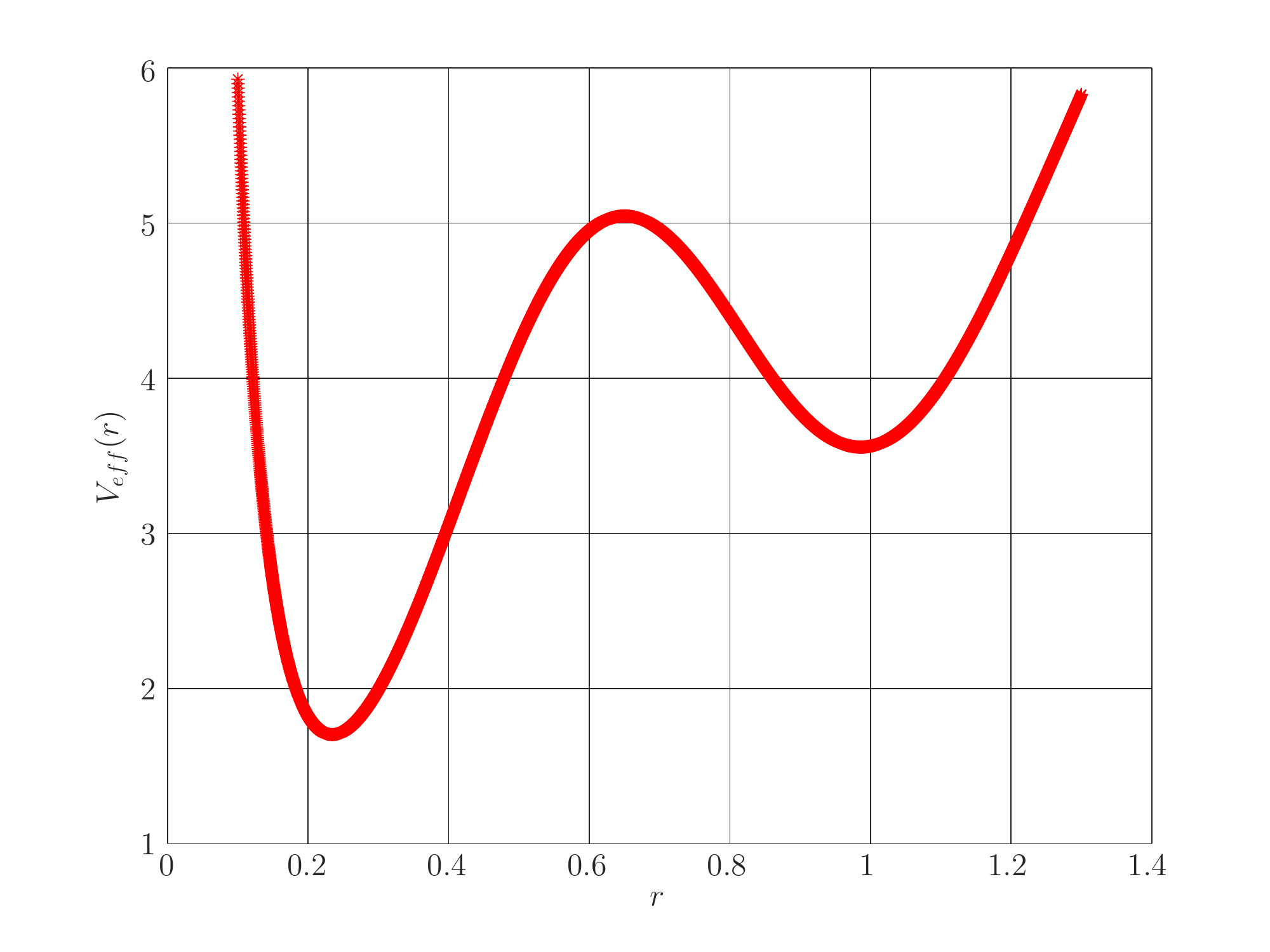}\par\end{centering}
\caption{(Color online) Effective potential for $F(r)=F_0 J_2(\lambda r)$. An Unstable point appears 
as expected around $r_0=0.6$. Here we choose the parameters as $F_{0}=1$, $p_\theta =0.5$, $p_z=1$, $B_0=4$. }
\label{fig:density-2} 
\end{figure}
Given
the expression (\ref{eq:Current_function_of_F}), and our choice of
$F$, expecting some current in $r=0$, implies that $\nu=2$, we
adjust $\lambda$ to keep a positive current up to $R=1$, which leads
to $\lambda=1/r_{1}$, with $r_{1}$ being the first zero of $J_{2}(r)$.
The shape of $J_{2}$, leads to expect a maximum near $r=1/2$ for
$F$, so we can expect that by tuning the parameters $a$ and $b$,
we will capture many $r_{0}$'s (for the different $p_{\theta}$'s),
giving rise to regions in phase space with separatices that have non-negligible
statistical weights. An illustration of possible obtained profiles
are depicted in Fig.~\eqref{fig:density-2}. 
\begin{figure}[t]
\begin{centering}
\includegraphics[width=8cm]{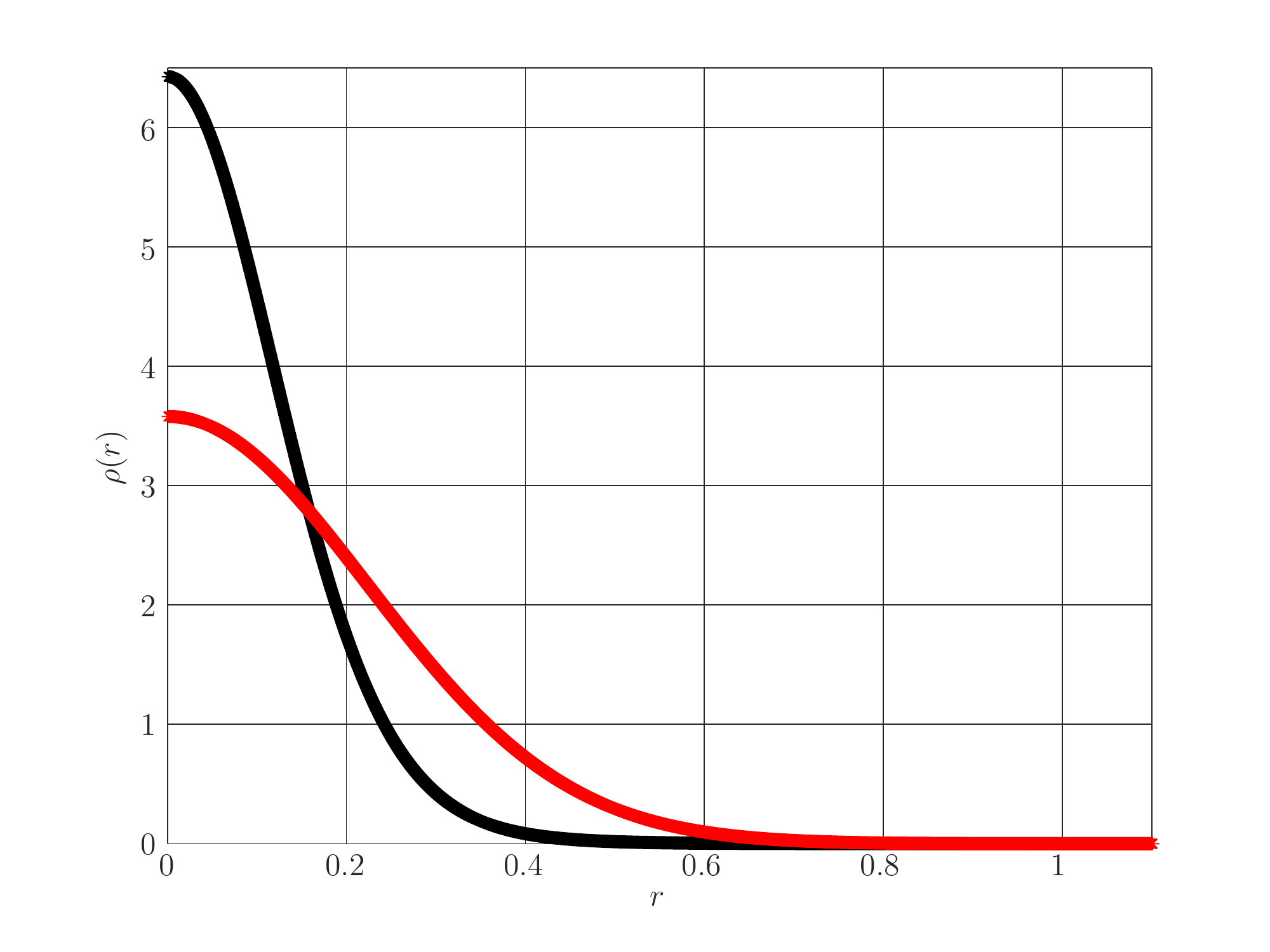} \\
\includegraphics[width=8cm]{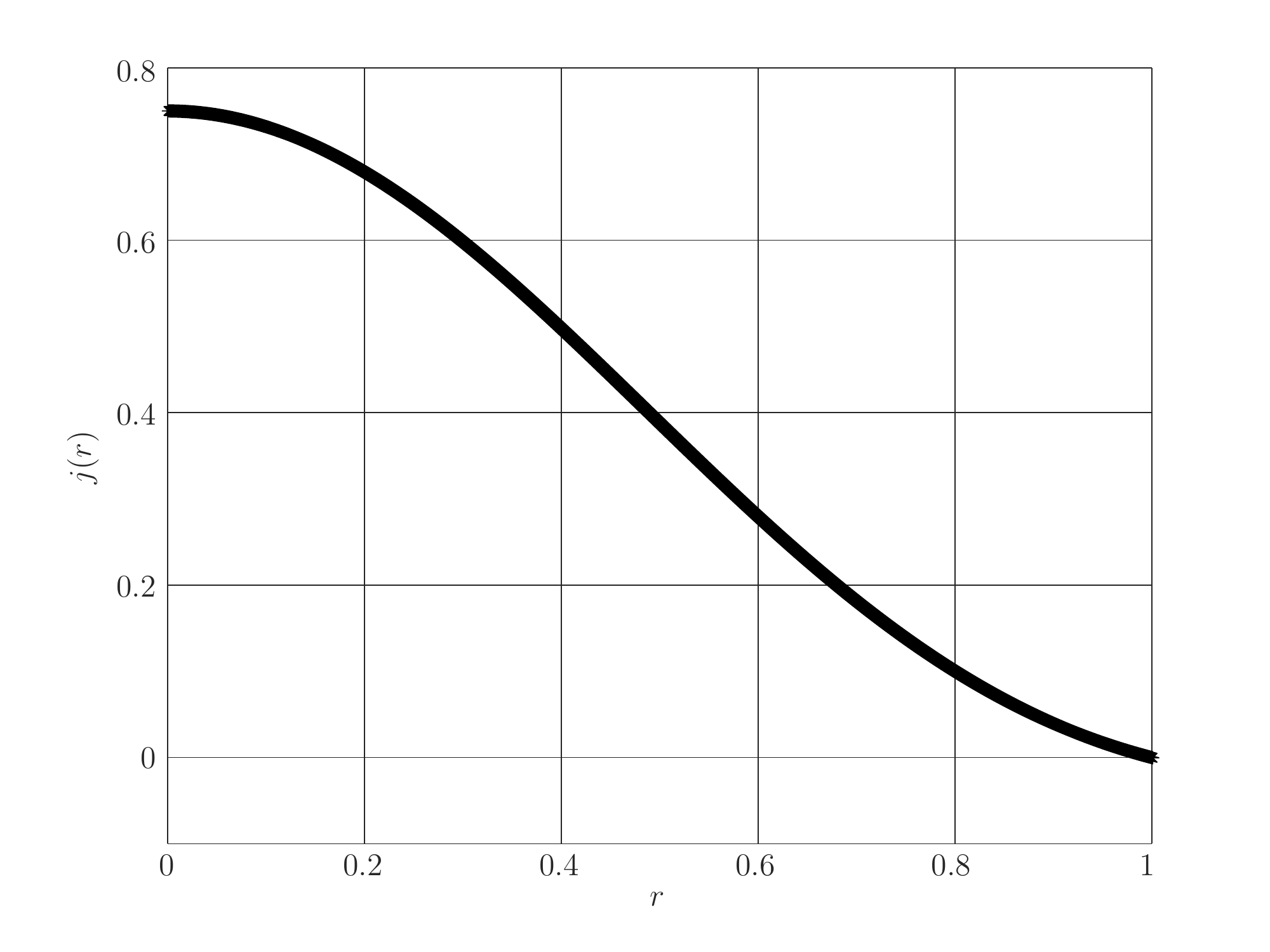}
\par\end{centering}
\caption{(Color online) Top: The gray(red) and black curves express the density
profiles with only one unstable point. Bottom: same current profile for
both configurations. The parameters are $F_{0}=1$, $a=b=10$ for
the gray(red) curve $F_{0}=1$, $a=10$, $b=0$ for the black. }
\label{fig:density-2} 
\end{figure}

Before concluding this letter we would like to make some remarks on
the observed profiles which indicate the presence of what we may call
an ITB although the underlying physical mechanisms are quite different.
We would like to stress out the similarities 
our observations and the magnetic ITBs discussed for instance in \cite{Constantinescu2012}.
One of the main difference is that if the plateau of $q(r)$ appears
the transport barrier emerges even if the value of $q$ is far from
rational number $m/n$ with small integers $m$ and $n$, in that sense
the creation of the barrier with the $q$-profile (\ref{eq:q}) is not  correlated to
the existence of a resonant surface, on the other hand for the considered
example, we can notice that the location of the magnetic ITB coincides with the place
on which the local density gradient exists.

Another study between the particle's motion and the existence of an
ITB using a pure field line approach and the existence of a stable
magnetic tori has been performed from the view point of the difference
between the magnetic winding number $q(r)$ and the effective one
$q_{{\rm eff}}(r)$ for the guiding venter orbit of the energetic
particles~\cite{Fiksel2005}. In a recent study~\cite{Ogawa2016-2},
it is shown that the resonance shift due to the grad $B$ drift and
its disappearance due to the curvature drift effect can create an
invariant tori in the particle dynamics while there are none for magnetic
field lines, and this is confirmed both analytically and numerically
with the full particle orbits around the resonance points. It should
be remarked that the guiding center theory is useless to clarify it
unlike Ref.~\cite{Ogawa2016-2}.

The above consideration provides de facto another difference between
the magnetic ITB and the effective ITB induced by the separatrices.
Moreover, we can stress out that the magnetic ITB is present in both
situations described in Fig.~\ref{fig:density}, while the steep
profile occurs only when the hyperbolic points are present. When
considering the degenerate q-profile we also note that the two unstable
fixed points appear around the magnetic ITB and when the parameter
$\lambda$ in Eq.~\eqref{eq:q} gets to be large, the steep region
of $\rho(r)$ gets to be strong as the influence of the separatrix
grows because the gap of $F(r)$ between $r<\alpha$ and $r>\alpha$
becomes smaller (see Fig.~\ref{fig:schematic6}).

To conclude, we have shown in this letter that steep equilibrium density
profiles can emerge due to the presence of a separatrix in the passive
particle orbits, this phenomenon is not related to the existence of
a local resonant surface and the observed phenomenon is reminiscent
of the presence of an ITB although the physical mechanisms inducing
it are a priori quite different in interpretation. We also discussed
how the $q$-profile can be tuned in order to generate such barriers.

We finally remark on what happens if we consider a toroidal configuration.
Let us imagine our cylindrical system is an infinite toroidal radius
limit of the toroidal system as Ref. \cite{Cambon2014}. The finite
toroidal radius effect breaks the integrability and it thus can induces
adiabatic chaos. It has been known for a while that the presence of
chaos affects sometimes the density profile locally. For instance,
the averaging effect in plasmas from the global chaos induced by the
resonance overlapping \cite{Chirikov1979} has been found and it modifies
the density profile \cite{White2012}. Even though we are directly
tackling the passive particle motions of ions and thus a different
type of localized chaos in this letter, similar things can be expected.
In the present case the unstable fixed points are located around the
place in which the steepness of density \eqref{eq:density} and the
local pressure gradient exist as shown in Fig.~\ref{fig:density}. Therefore the flattening effect makes them
steeper locally. As a result we can expect that this steepening effect
observed in the cylindrical configuration can be robust at least as
long as strongly chaotic motion remains localized near each hyperbolic
point.

\acknowledgments S. O. and X. L. thank G. Dif-Pradalier for useful
and encouraging discussions. X. L. thanks E. Laribi for careful reading of the manuscript. This work has been carried out within the framework of the French Research Federation for Magnetic Fusion
Studies. The project leading to this publication has received funding from Excellence Initiative of Aix-Marseille University - AMIDEX, a French ``Investissements d'Avenir'' programme.

\end{document}